\documentclass[%
 aip,
 amsmath,amssymb,
 preprint,%
]{revtex4-1}

\usepackage{graphicx}
\usepackage{dcolumn}
\usepackage{bm}

\usepackage{float}
\usepackage[utf8]{inputenc}
\usepackage[T1]{fontenc}
\usepackage{mathptmx}
\usepackage{etoolbox}
\usepackage{multirow}

\renewcommand{\Pr}{\operatorname{Pr}}
\renewcommand{\Re}{\operatorname{Re}}
\newcommand{\Nu}{\operatorname{Nu}}
\newcommand{\tdz}{{\theta'(0)}}

\makeatletter
\def\@email#1#2{%
 \endgroup
 \patchcmd{\titleblock@produce}
  {\frontmatter@RRAPformat}
  {\frontmatter@RRAPformat{\produce@RRAP{*#1\href{mailto:#2}{#2}}}\frontmatter@RRAPformat}
  {}{}
}%
\makeatother
\begin{document}

\preprint{AIP/123-QED}

\title{Exact solution for heat transfer across the Sakiadis boundary layer}
\author{W.\ Cade Reinberger}
\author{Nathaniel S.\ Barlow}%
\affiliation{ 
School of Mathematics and Statistics, Rochester Institute of Technology}%

\author{Mohamed A.\ Samaha}
\affiliation{%
Department of Mechanical and Industrial Engineering, Rochester Institute of Technology-Dubai
}%

\author{Steven J. Weinstein}
\affiliation{
Department of Chemical Engineering, Rochester Institute of Technology
}

\date{\today}

\allowdisplaybreaks

\begin{abstract}
We consider the problem of convective heat transfer across the laminar boundary-layer induced by an isothermal moving surface in a Newtonian fluid. In previous work  (Barlow, Reinberger, and Weinstein, 2024, \textit{Physics of Fluids}, \textbf{36} (031703), 1-3) an exact power series solution was provided for the hydrodynamic flow, often referred to as the Sakiadis boundary layer.  Here, we utilize this expression to develop an exact solution for the associated thermal boundary layer as characterized by the Prandtl number ($\Pr$) and local Reynolds number along the surface. To extract the location-dependent heat-transfer coefficient (expressed in dimensionless form as the Nusselt number), the dimensionless temperature gradient at the wall is required; this gradient is solely a function of $\Pr$, and is expressed as an integral of the exact boundary layer flow solution. We find that the exact solution for the temperature gradient is computationally unstable at large $\Pr$, and a large $\Pr$ expansion for the temperature gradient is obtained using Laplace's method.  A composite solution is obtained that is accurate to $O(10^{-10})$. Although divergent, the classical power series solution for the Sakiadis boundary layer---expanded about the wall---may be used to obtain all higher-order corrections in the asymptotic expansion. We show that this result is connected to the physics of large Prandtl number flows where the thickness of the hydrodynamic boundary layer is much larger than that of the thermal boundary layer. The present model is valid for all Prandtl numbers and attractive for ease of use.
\end{abstract}

\maketitle

\section{Introduction}
\label{intro}

The laminar boundary layer flow induced by a moving flat surface through an incompressible fluid has significant relevance to coating flows, and forms the basis for many studies involving moving elastic sheets with simultaneous heat and mass transport.  The boundary layer flow was first analyzed in 1961 by~\citet{Sakiadis1961a, Sakiadis1961b}, who utilized similarity variables to obtain a non-linear ordinary differential equation (ODE) for the dimensionless stream function identical to that of~\citet{Blasius}.  The Sakiadis solution, however, is distinctly different from that of Blasius.  In particular, the effect of a wall moving through an otherwise stationary fluid (Sakiadis) cannot be predicted via a simple translation of the solution for the flow induced by fluid moving across a stationary wall (Blasius).  ~\citet{Sakiadis1961b} solved the ODE system both numerically and via the Polhaussen approximation, in which a velocity profile is inserted into an integral form of the ODE system to obtain estimates of wall shear.  Since that time, there have been several numerical \cite{Cortell2010,Eftekhari2013,Fazio2015} and semi-analytical solutions \cite{Xu2013}  of the problem; in the past five years alone, the original Sakiadis boundary layer paper has been cited over 500 times which demonstrates its continued fundamental importance.  

In many applications, heat transfer accompanies the Sakiadis flow, and quality control of products is strongly dependent on the accurate estimation of the amount of heat transfer occurring at the moving surface.  As a result, there have been a number of experimental and theoretical studies to provide predictions of the Nusselt number (the dimensionless convective heat transfer coefficient) as a function of the dimensionless temperature gradient at the wall, which is itself dependent on the Prandtl number \cite{Tsou1967,Sound1980,Mout1980,Muc1979,Takhar1991,Pop1992,Pant2004,And2007,Girgin2011,Abd2017,Cortell2008,Daba2018,Abd1985,Hussaini1987,Afzal1993,Bianchi1993,Lin1994,Chen2000,Sparrow2005,Ishak2009,Bachok2012,Fang2003,Weidman2006}.  In these studies, the most accurate theoretical predictions are relegated to specific dimensionless Prandtl numbers since the underlying flow field is numerically determined (Table 1 below summarizes the $\Pr$ values examined in prior studies).  ~\citet{sak} obtained an exact power series solution for the Sakiadis flow which was recently made explicit in~\citet{exactPreprint}; here we use this solution to obtain an analytical expression for the temperature gradient at the wall for any Prandtl number.  Although exact, we find the solution is computationally unstable at moderate $\Pr$ values, and this motivates a large Pr expansion of the relevant integral of the flow field via Laplace’s method.  In the course of that analysis, we develop an efficient and generalizable implementation of Laplace’s method to obtain the asymptotic solution to all higher-order corrections.  We also prove that the standard divergent power series expansion about the wall may be used to obtain the large Pr asymptotic behavior exactly. Our results demonstrate explicitly that large $\Pr$ heat transfer calculations have a direct relationship to the shear stress along the moving wall.

\begin{figure}
\centering
\includegraphics[width=\textwidth]{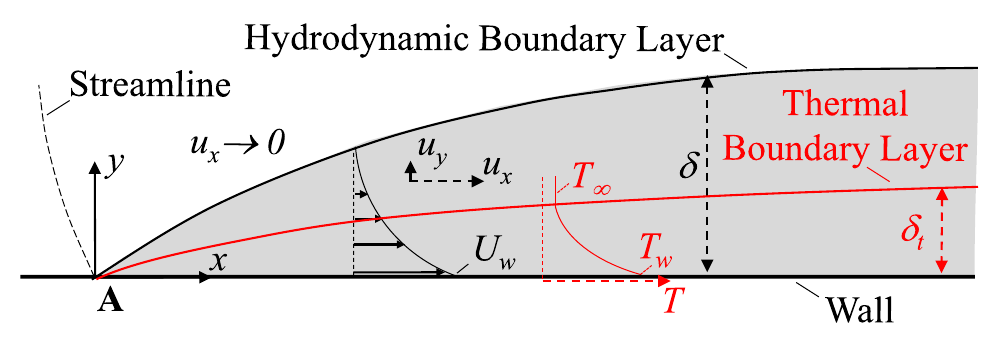}
\caption{Schematic of the development of both velocity and thermal boundary layers in Sakiadis flow.}
\label{SakBL}   
\end{figure}

We now provide a more precise description of the problem to solve, as well as necessary background for the current work. Consider two-dimensional flow of an incompressible Newtonian fluid with kinematic viscosity $\nu$, maintained in steady motion by a wall moving at velocity $U_w$ in the $x$-direction as shown in Figure \ref{SakBL}; the wall is fully responsible for the fluid motion. Owing to rapid speed of the wall, the velocity in the $x$-direction, $u_x$, is much larger than the velocity in the $y$-direction, $u_y$, and velocity gradients in the $y$ direction dominate those in the flow direction.  The effect of wall motion diminishes away from the wall, so $u_x \to 0$ as $y$ becomes large. The wall is maintained at a surface temperature, $T_w$, and the flowing fluid away from the wall has temperature $T_\infty$, such that convective heat transfer also occurs in the system.  As a result of the flow assumptions, the slope of the fluid streamlines is small, and the boundary layer approximation for fluid flow and heat transfer may be used.  Thus, the dominant momentum and energy transport is constrained to lie in small regions near the wall in the form of boundary layers in which the influence of the wall is constrained. Figure \ref{SakBL} shows the hydrodynamic and thermal boundary layers that form with respective thicknesses, $\delta$ and $\delta_t$.

In such a flow configuration, the velocity components may be related by the stream function $\psi$ and spatial variables may be contracted via a similarity transformation identical to that of Blasius \cite{Blasius}. In particular, the dimensionless stream function $f = \psi/\sqrt{\nu x U_w}$, may be expressed as a function of similarity coordinate $\eta = y\sqrt{U_w/\nu x}$. The $x$ and $y$ velocity components are subsequently expressed as $u_x = U_wf'$ and $u_y= \sqrt{\nu U_w /x} (\eta f' - f)/2$. Assigning a value of the stream function to be zero at the wall, the governing system for $f$---the Sakiadis boundary layer equations \cite{Sakiadis1961a}---are given as 

\begin{subequations}   
\label{mom}
\begin{equation} f'''(\eta)  + \frac{1}{2} f(\eta)f''(\eta)=0, \label{fode1} \end{equation} 
\begin{equation} f(0) = 0, f'(0)=1, f'(\infty)=0. \label{fbc1}\end{equation} 
 \end{subequations} Note that in (\ref{mom}) and throughout this paper, the prime symbols ($'$) denote derivatives with respect to the similarity variable $\eta$. The governing differential equation (\ref{fode1}), is identical to that of Blasius, although the location of the derivative boundary conditions in (\ref{fbc1}) are swapped. 

For cases where the fluid and transport properties are approximated as temperature independent, the convective heat transfer may be solved subsequently to the fluid flow, and the same similarity transformation as for flow may be used. The dimensionless temperature, $\theta(\eta) = (T - T_w)/(T_\infty - T_w)$, is thus governed by the well-known energy equation \cite{Schlichting, Pohlhausen1921} given as 

\begin{subequations}  \label{PohlSys} \begin{align} \theta''(\eta) & + \frac{1}{2}\Pr f(\eta)\theta'(\eta)=0, \label{PohlODE}  \\ \theta(0) &= 0, \hphantom{\sum\sum} \theta(\infty)=1. \label{bc1} \end{align}  \label{firstbruh} \end{subequations} As evidenced from (\ref{PohlODE}), $\tdz$ is a function of the Prandtl number, $\Pr=\nu/\alpha$, where $\alpha$ is the thermal diffusivity. Note the system (\ref{PohlSys}) is linear, and may be solved by quadrature in a straightforward way.

The work of~\citet{sak} followed by~\citet{exactPreprint} provide an exact analytic expression for $f(\eta)$ given as \begin{subequations} \label{at} \begin{align}  f(\eta) &= \sum_{n=0}^\infty \tilde{a}_n e^{-nC\eta/2} \hphantom{\sum} \eta \in [0, \infty), \\ \tilde{a}_{n+1} &= \frac{1}{Cn(n+1)^2}\sum_{k=1}^n k^2 \tilde{a}_k \tilde{a}_{n+1-k}, n \geq 1, \label{rec} \\ \nonumber C &= \tilde{a}_0 =  1.6161254468..., \\ \tilde{a}_1 &= -2.131345924047...\end{align} \label{welpthismyman}\end{subequations}
In (\ref{welpthismyman}) the quantities $C = \tilde{a_0}$ and $\tilde{a}_1$, terminated above at double precision, were calculated by~\citet{sak} using finite truncations of (\ref{welpthismyman}) coupled with an iterative technique that enforces the conditions at $\eta=0$ in (\ref{fbc1}); the recursion (\ref{rec}) was independently obtained by~\citet{KirsurJoshi} for the related problem of a moving wedge,  where an iterative technique was also employed.  More recently \cite{exactPreprint}, exact power series representations of these quantities have been determined, which also enable an explicit analytical expression for the dimensionless wall shear stress parameter, $\kappa$; these expressions are summarized below.

\begin{subequations}
    \begin{align}
        C = \tilde{a}_0 &= \sqrt{\displaystyle -\frac{2}{\zeta} \sum_{n=1}^\infty n b_n (-1)^{n-1}}, \hphantom{\sum} \tilde{a}_1 = \frac{C}{\zeta}, \\ 
        \kappa &= -\frac{C}{2} + \dfrac{2}{C\zeta^3} \sum_{n=2}^\infty n(n-1)b_n (-1)^n,  \\ 
        \zeta &= \sum_{n=1}^\infty b_n (-1)^n, \hphantom{\sum} b_{n+1} = \frac{D_{n, n+1}}{n+1} \hphantom{\sum} (n \geq 1), \hphantom{\sum} b_1 = 1, \\ 
        D_{1, m} &= -\frac{m}{4} \hphantom{sum} (m \geq 1),  \\  D_{n, m} &= -mA_{n+1} - \frac{1}{n} \sum_{j=1}^{n-1} (jm+n-j)A_{j+1} \hphantom{\sum} (n \geq 1, m \geq 2), \\
        A_{n+1} &= \frac{1}{n(n+1)^2} \sum_{j=1}^n j^2 A_j A_{n-j+1} \hphantom{\sum} (n \geq 1),  \hphantom{\sum} A_0 = A_1 = 1.
    \end{align} \label{exKappa}
\end{subequations} The parameter $\kappa$ is an essential physical result often extracted from flow studies.

The solution of (\ref{firstbruh}) for the dimensionless temperature field leads to an expression of the local Nusselt number as a function of the temperature gradient at the wall \cite{Tsou1967}, $\tdz$, as

\begin{equation} \Nu_x = \tdz \Re_x^{1/2}. \label{nudef} \end{equation}  

\noindent In (\ref{nudef}), $\Nu_x = hk/x$, where $h$ and $k$ are the dimensional convective heat transfer coefficient and thermal conductivity of the flowing fluid, respectively, and $\Re_x = xU_w/\nu$. Table \ref{PrevStud} shows a summary of prior work with predicted values of $\tdz$. Also shown in the table are values of $\kappa$ in those studies.  Note that the velocity field is used as an input to calculate $\tdz$, and thus the accuracy in $\kappa$ restricts the accuracy in $\tdz$. Prior work also indicates that, as stated above, numerical simulations have been relegated to specific Prandtl numbers, and interpolation is needed to extract $\tdz$ for intermediate Prandtl numbers. In our work, we will provide a solution for $\tdz$ that leads to a model valid for any $\Pr$, and with higher accuracy than those values shown in Table \ref{PrevStud}.

\begin{table}[H]
\caption{ Previous studies and their reported results of dimensionless wall shear stress parameter, $\kappa = f''(0)$, and the dimensionless temperature gradient at the wall $\tdz$, for various Prandtl numbers, $\Pr$.}

\begin{center}

\scalebox{.8}{

\label{PrevStud}
\begin{tabular}{|l|l|l|l|l|}

\hline

\hline

{\bf Author(s)}	& {\bf Year} 	&	$ \boldsymbol{-\kappa} $	&	\textbf{Pr}		&	$ \boldsymbol{\theta}{\bf '(0)} $	 \\

\hline

\hline

Sakiadis \cite{Sakiadis1961b} 	&	1961 &   	 		0.44375 & 	&  \\

\hline

\multirow{4}{*}{Tsou et al.\ \cite{Tsou1967}}	& \multirow{4}{*}{1967}	&		 \multirow{4}{*}{0.444} & 0.7& 	0.3492219		 \\

\cline{4-5}

   &  	& &	1	 & 	0.4438		 \\

\cline{4-5}

   & 	& &	10	 & 	1.68043435		 \\

\cline{4-5}

 & 	& &	100	 & 	5.545		 \\

\hline

\multirow{3}{*}{Soundalgekar and Murty \cite{Sound1980}}	&	\multirow{3}{*}{1980} &   \multirow{3}{*}{}	& 	0.7	 & 0.3508	 \\
\cline{4-5}

   &  	& &	2	 & 	0.6831		 \\
\cline{4-5}

   & 	& &	10	 & 	1.6808		 \\
\hline

\multirow{2}{*}{Moutsoglou et al.\ \cite{Mout1980}}	& \multirow{2}{*}{1980}	&		 \multirow{2}{*}{0.44375} & 0.7& 	0.34924		 \\

\cline{4-5}

   &  	& &	7	 & 	1.38703		 \\
\hline

Takhar et al. \cite{Takhar1991} 	&	1991 &   	 		0.4439 &  0.7	& 0.3508 \\

\hline

Pop et al. \cite{Pop1992} 	&	1992 &   	 		0.4445517 &  0.7	& 0.3507366 \\

\hline

Pantokratoras \cite{Pant2004} 	&	2004 &   	 		0.4438 &  0.7	& 0.35 \\

\hline

\multirow{2}{*}{Andersson and Aarseth \cite{And2007}}	& \multirow{2}{*}{2007}	&		 \multirow{2}{*}{0.4437483} & 0.7& 	0.3492359		 \\

\cline{4-5}

   &  	& &	10	 & 	1.680293		 \\

\hline

\multirow{6}{*}{Cortell \cite{Cortell2008}}	& \multirow{6}{*}{2008}	&		 \multirow{6}{*}{0.44374733} & 0.6& 	0.3135188		 \\

\cline{4-5}

   &  	& &	5.5	 & 	1.216049		 \\

\cline{4-5}

   & 	& &	7	 & 	1.387033		 \\

\cline{4-5}

 & 	& &	10	 & 	1.680293		 \\

\cline{4-5}

 & 	& &	50	 & 	3.890918	 \\

\cline{4-5}

 & 	& &	100	 & 	5.544663		 \\

\hline

\multirow{2}{*}{Girgin \cite{Girgin2011}}	& \multirow{2}{*}{2011}	&		 \multirow{2}{*}{0.4437483123} & 0.7& 	0.3492358481		 \\

\cline{4-5}

   &  	& &	10	 & 	1.6802932833		 \\

\hline

\multirow{3}{*}{Bachok et al.\ \cite{Bachok2012}}	&	\multirow{3}{*}{2012} &   \multirow{3}{*}{0.4437}	& 	0.7	 &0.3492	 \\
\cline{4-5}

   &  	& &	1	 & 	0.4437		 \\
\cline{4-5}

   & 	& &	10	 & 	1.6803		 \\

\hline

Abdella et al.\ \cite{Abd2017} 	&	2017 &   	 		0.4426557 &  0.7	& 0.3493033 \\

\hline

Barlow et al.\ \cite{Nate2017} 	&	2017 &   	 		0.443748313369 & 	&  \\

\hline

\multirow{5}{*}{Daba and Tuge \cite{Daba2018}}	& \multirow{5}{*}{2018}	&		 \multirow{5}{*}{0.4437487181} & 0.7& 	0.349241899		 \\

\cline{4-5}

   &  	& &	1	 & 	0.443748718		 \\

\cline{4-5}

 & 	& &	10	 & 	1.68032745	 \\

\cline{4-5}

 & 	& &	100	 & 	5.546202431	 \\


\hline

Naghshineh et al.\ \cite{sak} 	&	2023 &   	 		0.443748313368861 & 	&  \\

\hline 

Barlow et al.\ \cite{exactPreprint} & 2024 & \text{Exact} & & \\

\hline

\hline

\end{tabular}
}

\end{center}
\end{table}

The organization of this paper is as follows.  Section~\ref{Exact} provides both the exact and large Pr analytical solutions for the heat transfer problem.  Section~\ref{ApproxMod} explains the assumptions and analysis of the approximate model using the physics of large Prandtl number flows. Results, discussions and validation of the presented models are compared to prior studies in Section~\ref{Results}, followed by conclusions in Section~\ref{Conc}. Appendices are provided to provide more detailed mathematical analyses that support the presented work.

\section{Analytical model}
\label{Exact}

Two complimentary expressions for the temperature gradient at the wall, i.e. $\tdz$ in (\ref{nudef}), are obtained, which are taken together to provide the analytic representation of the solution. We detail the development of these expansions as follows. In Section 2.1, we develop an exact series solution for the temperature gradient. The series, however, becomes poorly conditioned for $\Pr > 12$ in double arithmetic due to finite precision limitations. We thus develop, in Section 2.2, the asymptotic expansion of the temperature gradient for large $\Pr$ by employing Laplace's method on a relevant integral that arises in the analysis. A fully analytical composite solution is developed in Section 2.3 that weaves together the exact (Section 2.1)  and asymptotic (Section 2.2) expressions.

\subsection{Exact analytical expression for $\tdz$}
\label{smallPr} 

To proceed, we substitute the exact solution to the system (\ref{mom}) for $f$, given by (\ref{welpthismyman}), into the ODE system (\ref{firstbruh}). Next, we rewrite our differential equation as a logarithmic derivative, solve for $\theta'$, and integrate to obtain
\begin{multline} \frac{\theta''}{\theta'} =\left(\ln(\theta')\right)' = -\frac{1}{2}\Pr f \\ \Rightarrow  \theta'(\eta) = H \exp\left(-\frac{1}{2}\Pr \int_0^\eta f(\xi) \ \text{d}\xi \right) \\ \Rightarrow \theta(\eta) = H \int_\eta^\infty \exp\left(-\frac{1}{2}\Pr \int_0^\eta f(\xi) \ \text{d}\xi\right) \ \text{d}\eta + F. \label{tdzgen} \end{multline}
In (\ref{tdzgen}) $F$ and $H$ are integration constants. The boundary condition at $\eta \to \infty$ in (\ref{bc1}) indicates that $F=1$ in (\ref{tdzgen}), and subsequently applying boundary condition (\ref{bc1}) yields \begin{equation} \tdz = H = \left\{\int_0^\infty \exp\left(-\frac{1}{2}\Pr \int_0^\eta f(\xi) \ \text{d}\xi\right) \ \text{d}\eta\right\}^{-1}, \label{genfin} \end{equation}
which provides an explicit expression for $\tdz$, and thereby Nusselt number via (\ref{nudef}), in terms of the dimensionless stream function, $f$; the solution for $f$ satisfies the system (\ref{mom}) and is given by the analytical solution (\ref{welpthismyman}). 

 Upon substitution of the expansion (\ref{at}) into (\ref{genfin}), we obtain $$\int_0^\eta f(\xi) \ \text{d}\xi = C\eta + 2 \sum_{n=1}^\infty \left\{\frac{\tilde{a}_n}{nC}\left(1 -e^{-nC\eta/2}\right)\right\}, $$ so that \begin{multline*} \tdz = \left\{\int_0^\infty \exp\left(-\frac{C}{2}\Pr \eta \vphantom{\int_0^\infty \exp\left(-\frac{C}{2}\Pr \eta + \Pr \sum_{n=1}^\infty \frac{\tilde{a}_n}{nC} e^{-nC\eta/2} \right) \ \text{d}\eta} \right. \right. \\ \left. \left. \vphantom{\int_0^\infty \exp\left(-\frac{C}{2}\Pr \eta + \Pr \sum_{n=1}^\infty \frac{\tilde{a}_n}{nC} e^{-nC\eta/2} \right) \ \text{d}\eta}+ \Pr \sum_{n=1}^\infty \frac{\tilde{a}_n}{nC} e^{-nC\eta/2} \right) \ \text{d}\eta \right\}^{-1} \exp\left(\Pr \sum_{n=1}^\infty \frac{\tilde{a}_n}{nC} \right). \end{multline*} To proceed further, we change the integration variable to $u = \exp(-C\eta/2)$, to yield \begin{multline} \tdz = \left\{\frac{2}{C} \int_0^1 u^{\Pr-1} \exp\left(\Pr \sum_{n=1}^\infty \frac{\tilde{a}_n}{nC}u^n \right) \ \text{d}u \right\}^{-1} \\ \times ~~~ \exp\left(\Pr \sum_{n=1}^\infty \frac{\tilde{a}_n}{nC}\right). \label{keyint} \end{multline} 
The coefficients $\tilde{a}_n$ decay exponentially in magnitude, and we find that the sum inside the rightmost exponential in (\ref{keyint}) converges to double machine precision in 35 terms. Thus, the only remaining issue is the computation of the integral in (\ref{keyint}). This evaluation may be achieved by making use of a recursive formula for the exponential of a series \cite{Gibbons}. In particular, we write \begin{align} \sum_{n=0}^\infty b_n u^n &= \exp\left(\Pr\sum_{n=1}^\infty \frac{\tilde{a}_n}{nC} u^n\right),  \label{expeval} \end{align} \begin{subequations}  \begin{align}  b_{n+1} &= \frac{\Pr}{C(n+1)}\sum_{k=0}^n \tilde{a}_{k+1}b_{n-k}, b_0 = 1. \label{brec} \end{align} \label{mygen} 
Inserting (\ref{mygen}) into (\ref{keyint}) and integrating term-by-term we obtain \begin{equation} \tdz = \frac{C}{2}\exp\left(\Pr\sum_{n=1}^\infty \frac{\tilde{a}_n}{nC}\right)\left\{\sum_{n=0}^\infty \frac{b_n}{n+\Pr}\right\}^{-1}. \label{sergen}\end{equation} \end{subequations}
The result, (\ref{sergen}), is an exact and pointwise convergent representation of $\tdz$ that can be substituted in (\ref{nudef}) to provide the exact value of Nusselt number. Although an exact result, the recurrence relation for $b_n$, (\ref{brec}), becomes poorly conditioned for large Prandtl number, and an alternative methodology is discussed in Section 2.2. to overcome this deficiency.  We will discuss the limitations of the solution (\ref{sergen}) in Section 2.3.

\subsection{$\tdz$ for Large $\Pr$}
\label{LargePr} 

As mentioned above, the series solution of (\ref{sergen}) for large $\Pr$ has computational limitations. Therefore, we develop an asymptotic expansion for $\tdz$ as $\Pr \to \infty.$  To do so, we utilize Laplace's method to approximate the integral in (\ref{keyint}) and include higher-order corrections. The details of the analysis are provided in Appendix~\ref{AppA}, but the key elements are now discussed. We note that the integral in the denominator of (\ref{keyint}), denoted here as $I$, is of the form \begin{subequations} \begin{align}  I &= \int_0^1 \frac{1}{u}\exp(\phi(u)\Pr) \ \text{d}u, \label{intlapform} \\ \phi(u) &= \ln(u) + \sum_{n=1}^\infty \frac{\tilde{a}_n}{nC}u^n.\label{phidef}\end{align}\label{jjettasiguessimouttajokes}\end{subequations}
Over the domain of the integral, we find that $\phi(u)$ has its maximum at $u=1$, and $\phi'(u) = 0$ at that location. In accordance with Laplace's method, $\phi(u)$ and $u^{-1}$ in the integrand of (\ref{intlapform}) are expanded in a Taylor series about $u=1$, and the lower limit of integration is extended to minus infinity. The latter step is afforded because the dominant contribution of the integral occurs in the vicinity of $u=1$, and owing to the exponential term in the integrand, subdominant errors are incurred with this domain extension. After organizing terms at each order of $\Pr$, one arrives at the expansion \begin{equation} I \sim \frac{C}{2}e^{\phi(1)\Pr}\left\{\frac{\gamma_1}{\Pr^{1/2}} + \frac{\gamma_2}{\Pr} + \frac{\gamma_3}{\Pr^{3/2}} + \frac{\gamma_4}{\Pr^2} + \dots \right\} \hphantom{\sum} (\Pr \to \infty). \label{intasym} \end{equation} The coefficients $\gamma_n$ in (\ref{intasym}) are provided in Appendix~\ref{GamTab}, which are expressed solely in terms of the shear stress parameter at the wall, $\kappa$. It is remarkable that the constant $C$, associated with the $\eta \to \infty$ asymptotic behavior of the function $f$ (see (\ref{welpthismyman})), does not appear in the final expressions for the coefficients $\gamma_n$---that is despite its explicit appearance in equation (\ref{phidef}).  An explanation for this observation is provided in Section 3.2.

Using the expansion (\ref{intasym}) in the solution (\ref{keyint}), we obtain \begin{equation}  \tdz^{-1} \sim \sum_{n=1}^\infty \gamma_n \Pr^{-n/2} \label{gexp} \hphantom{\sum} (\Pr \to \infty).\end{equation} The asymptotic series (\ref{gexp}) is divergent; that is, for fixed $\Pr$, including additional terms eventually leads to series divergence. As is common with asymptotic series, however, (\ref{gexp}) has convergent character for a few terms at a fixed $\Pr$, beyond which divergence is observed. The highest accuracy of the series occurs at its optimal truncation---this is the number of terms just before the series starts diverging \cite{BenderAndOrszag}. The expansion can be further simplified using the long-known recursion for the reciprocal of a power series \cite{Euler, Hansted} to yield an explicit asymptotic expansion for $\tdz$ as a function of $\Pr$ as \begin{subequations} \begin{align} \tdz &\sim  \Pr^{1/2}\sum_{n=0}^\infty \delta_n \Pr^{-n/2}, \\ \delta_n &= \frac{-1}{\gamma_1}\sum_{k=1}^n \gamma_{k+1}\delta_{n-k}, \text{ with } \delta_0 = \gamma_1^{-1}. \end{align} \label{dexp} \end{subequations}. 

\subsection{Composite hybrid solution for $\tdz$}
\label{comp}

\begin{figure}
    \centering
    \includegraphics[width=\textwidth]{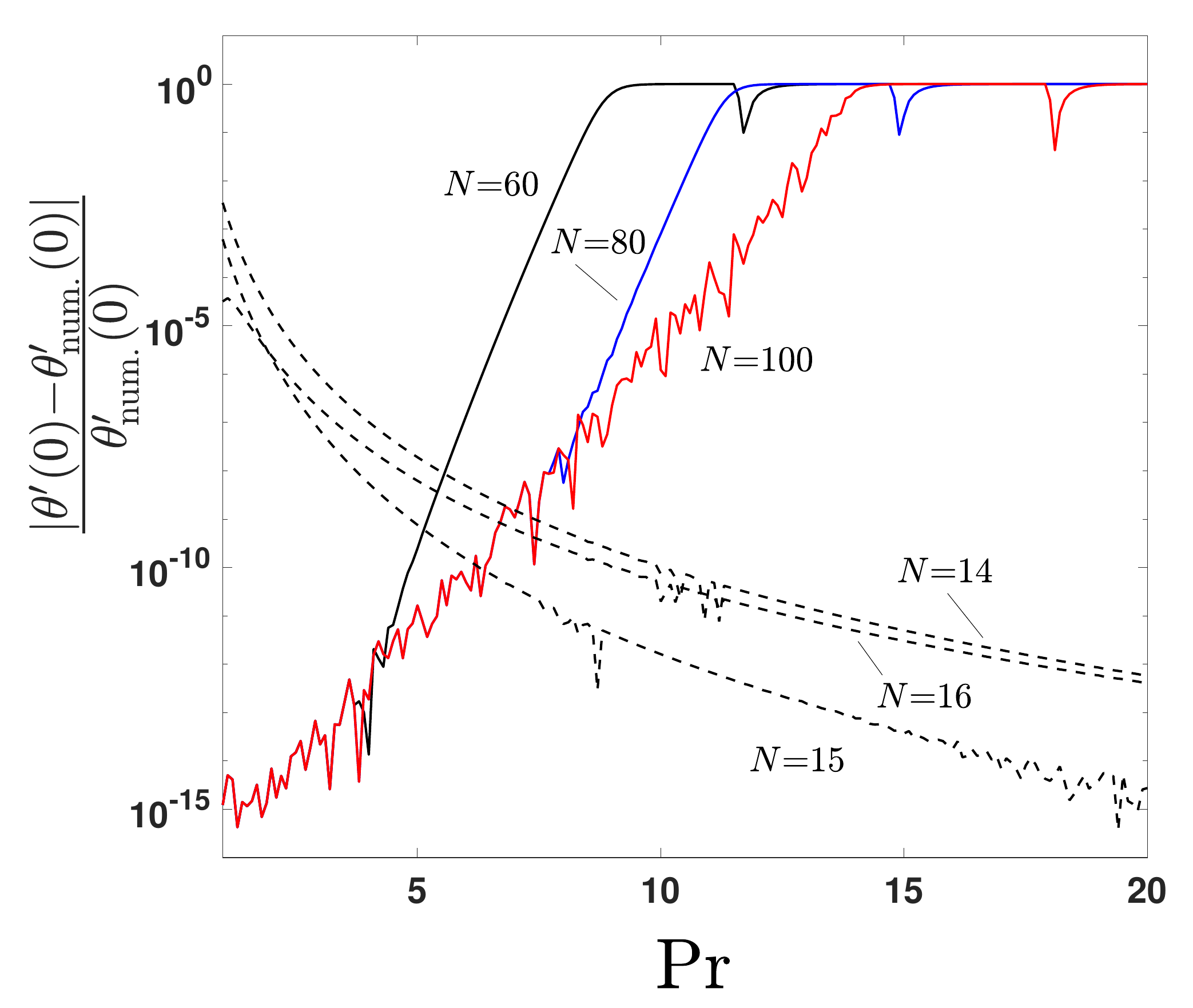}
    \caption{Error of analytical solutions relative to the numerical solution of (\ref{keyint}). The solid curves indicate truncations of expansion~(\ref{sergen}) (the exact solution), computed in double precision, while the dashed curves represent truncations of the asymptotic expansion~(\ref{dexp}). In both cases, $N$ denotes the order of truncation. For certain values of $\Pr$, the error increases between $15$ and $16$ terms of the asymptotic solution, indicating that there is an optimal truncation.}
    \label{fig:crossover}
\end{figure}

We have presented two expressions for $\tdz$ as a function of $\Pr$, needed to express the dimensionless convective heat transfer coefficient at the solid moving wall according to (\ref{nudef}). The first expression is a pointwise convergent and exact series solution for small $\Pr$ given by (\ref{sergen}). The second expression is an asymptotic expansion in powers of $\Pr^{-1/2}$ for large $\Pr$, given by (\ref{dexp}), which is useful up to its optimal truncation for a given $\Pr$. We aim to find the optimal truncation of the intersection of the asymptotic expansion with the \textit{curve} given by (\ref{sergen}) taken to its computational limit. We thus combine these two expressions in a way that minimizes the error for all $\Pr$. To do so, we plot the error of each expansion relative to the numerical evaluation using Simpson's rule of the integral in (\ref{keyint}) with an approximate error tolerance of $10^{-11}$ as shown in Figure \ref{fig:crossover}. 

For a fixed series truncation $N$, it is seen that (\ref{sergen}) yields an error curve that increases with $\Pr$ due to the computational errors referred to above. Meanwhile, the asymptotic expansion (\ref{dexp}) produces error curves for fixed number of terms that decrease with $\Pr$. As the number of terms in the asymptotic expansion (\ref{dexp}) increase, the values of $\Pr$ for which the expansion remains accurate moves to larger $\Pr$---this is a direct consequence of optimal truncation.  As shown in Figure \ref{fig:crossover}, we see that $80$ terms of (\ref{sergen}) and $15$ terms of (\ref{dexp}) give a global maximum error or order $10^{-10}$. 

The resulting final optimal expression for $\tdz$ is thus given as 
\begin{subequations}
\begin{align}
    \tdz &= \begin{cases} \displaystyle \frac{C}{2}\exp\left(\Pr\sum_{n=1}^{80} \frac{\tilde{a}_n}{nC}\right) \left\{\sum_{n=0}^{80} \frac{b_n}{n+\Pr}\right\}^{-1} & \Pr < 6 \\  \displaystyle \tdz \sim  \Pr^{1/2}\sum_{n=0}^{15} \delta_n \Pr^{-n/2} & \Pr \geq 6  \end{cases}, \\ \delta_n &= \frac{-1}{\gamma_1}\sum_{k=1}^n \gamma_{k+1}\delta_{n-k},~~~\delta_0 = \frac{1}{\gamma_1}, \end{align} \centering{($\gamma_n$  given in Appendix~\ref{GamTab}),} \begin{align} b_{n+1} &= \frac{\Pr}{C(n+1)}\sum_{k=0}^n \tilde{a}_{k+1}b_{n-k}, \hphantom{\sum} b_0 = 1,  \\ \tilde{a}_{n+1} &= \frac{1}{Cn(n+1)^2}\sum_{k=1}^n k^2 \tilde{a}_k \tilde{a}_{n+1-k}, \\ C &= \tilde{a}_0 = 1.6161254468... \hphantom{\sum} \tilde{a}_1 = -2.131345924047.... \label{constswelpa}
\end{align} \label{thebigsum}
\end{subequations}\noindent where exact expressions for $C=\tilde{a}_0$ and $\tilde{a}_1$ are given by (\ref{exKappa}), if one requires more digits than provided in (\ref{constswelpa}). 

\section{The equivalence of Laplace's Method for $\Pr\to\infty$ and approximations employing polynomial velocity profiles} \label{ApproxMod}

Significant physical insights and mathematical simplifications may be obtained by considering an engineering approach to solving for $\tdz$ for large $\Pr$.  In Section 3.1, we follow the Polhausen approach of assuming a polynomial profile for the dimensionless stream function $f$ in the evaluation of $\tdz$ in (\ref{keyint}).  We examine the implications of these results to the evaluation of $\tdz$ in Section 3.2 via Laplace's method, and propose an alternative simple method to do so.

\subsection{Determining $\tdz$ via low order polynomial expression for velocity}

At large $\Pr$, the thickness of the thermal boundary layer ($\delta_t$) is thinner than that of the hydrodynamic boundary layer ($\delta$); see Figure 1 for a definition sketch. Consequently, the thermal boundary layer is confined in a zone where the velocity, $u_x=U_wf'(\eta)$, is linear with the dimensionless distance from the wall $\eta$. A similar assumption was first introduced by~\citet{Lev1928} for the Blasius boundary-layer flow and can be considered as analogous to the polynomial profiles used in the Polhausen approximation for the wall shear (although as we will show, the boundary conditions used in the classical Polhausen approximation must be modified from analogy with the Polhausen approximation in order to obtain the correct asymptotics for large $\Pr$ flows). 


The stream function $f$ is parabolic for a linear velocity approximation; the constraints in (\ref{fbc1}) at $\eta = 0$ as well as $f''(0) = \kappa$ are applied to yield \begin{subequations}\begin{equation} \label{fstr} f=\kappa\dfrac{\eta^2}{2}+\eta, \hphantom{\sum} \text{ for } \eta \leq \eta_t \end{equation} where $\eta_t$ is a representation of the nondimensionalized distance to the edge of the boundary. Note that the locus of points, $\eta=\eta_t$, maps out the location of the thermal boundary layer, $\delta_t$, when mapped back to physical $x$-$y$ coordinates, and is given by \begin{equation} y=\delta_t=\eta_t\sqrt{\dfrac{\nu x}{U_w}}. \end{equation} \end{subequations} Substituting (\ref{fstr}) in the Pohlhausen ODE, (\ref{PohlODE}), gives \begin{equation} \label{PohlLinVel} \dfrac{\theta''}{\theta'}=-\Pr\left(\kappa\dfrac{\eta^2}{4}+\dfrac{\eta}{2}\right), \hphantom{\sum} \text{ for } \eta < \eta_t. \end{equation} Integrating (\ref{PohlLinVel}) and employing the boundary conditions, Eqs.~(\ref{bc1}), gives \begin{subequations} \begin{equation} \label{thetaD1} \theta'=\theta'(0) \exp \left(-\Pr\left(\kappa\dfrac{\eta^3}{12}+\dfrac{\eta^2}{4}\right)\right), \hphantom{\sum} \text{ for } \eta < \eta_t.  \end{equation}

Note that $\tdz$ is an unknown quantity at this point, so is in fact an integration constant.  At the edge of the thermal boundary layer ($\eta=\eta_t$), we expect $\theta'$ to be zero.  However, the functional form of (\ref{thetaD1}) does not allow that choice.  Instead, we choose the minimum of $\theta'$ to be zero at the edge of the boundary layer, i.e. $\theta''=0$, and thus from (\ref{PohlLinVel}) we have \begin{equation} \label{valid} \eta_t=-\dfrac{2}{\kappa}. \end{equation} \end{subequations} In order to find $\tdz$, (\ref{thetaD1}) is integrated across the boundary layer to yield \begin{equation} \label{theta11} \theta(\eta_t)-\theta(0)=\theta'(0) \int_0^{\eta_t} \exp \left(-\Pr\left(\kappa\dfrac{\eta^3}{12}+\dfrac{\eta^2}{4}\right)\right)\text{d}\eta. \end{equation} Consistent with the approximations made thus far, we assume that $\theta(\eta_t) = 1$, which is the infinite boundary condition in (\ref{bc1}) located at the boundary layer edge. Using this constraint we obtain \begin{equation} \label{thetaD111} \tdz= \left[\int_0^{\eta_t} \exp \left(-\Pr\left(\kappa\dfrac{\eta^3}{12}+\dfrac{\eta^2}{4}\right)\right)\text{d}\eta\right]^{-1}. \end{equation}

Let us now summarize the approximations embedded in (\ref{thetaD111}). First, we have assumed a velocity profile to obtain $f$.  Second, we used the fact that $\theta''$ is zero at the edge of the boundary layer, but $\theta'(\eta) \neq 0$ there.  Finally, we have displaced the boundary condition on theta from infinity down to $\eta=\eta_t$.  Considering the approximations made at the edge of the boundary layer, there is no reason to evaluate (\ref{thetaD111}) exactly.  Instead, we utilize Laplace's method on (\ref{thetaD111}), which uses the fact that the maximum of the argument of the exponential occurs at $\eta=0$, and this dominates the behavior of the integral as $\Pr\to\infty$. An intuitive advantage of this method is that the integral is ``weighted" away from the most approximated location at $\eta=\eta_t$.  Laplace's method may be implemented by expanding $$\exp \left(-\Pr\left(\kappa\dfrac{\eta^3}{12}\right)\right) = \displaystyle\sum_{n=0}^\infty \dfrac{(-\kappa/12)^n}{n!}\Pr^n \eta^{3n},$$ in (\ref{thetaD111}) and moving the upper limit of the integral to infinity; this step only incurs asymptotically small errors. The resulting expression $\tdz$ can be expressed in terms of particular values of the Gamma function as \begin{equation} \label{thetaD1111} \tdz\approx \left[\displaystyle\sum_{n=0}^\infty\dfrac{\Gamma\left(\dfrac{3n+1}{2}\right)}{n!} \left(\dfrac{-2\kappa}{3}\right)^n\Pr^{-(n+1)/2}\right]^{-1}.
\end{equation}

Remarkably, the first three terms of (\ref{thetaD1111}) are identical to the correct asymptotic expansion of $\tdz^{-1}$ in the $\Pr \to \infty$ limit given by (\ref{dexp}). Recall that this result utilized the exact solution for $f$ coupled with Laplace's method.   In fact, in the next Section, we show that the full Laplace method expansion may be obtained via a mathematical approach motivated by this observation.

\subsection{Full Laplace Expansion as $\Pr \to \infty$ using Divergent Infinite Series Solution}

The fact that a low-order polynomial approximation was used to recover the first few terms in the asymptotic expansion (\ref{gexp}) motivates the examination of higher-order polynomials to generate higher-order corrections.  We now discuss such an analysis.  In Appendix A.1, Laplace’s method is used to determine the large $\Pr$ behavior of the integral (\ref{keyint}); this integral is embedded in the desired Nusselt number expression (\ref{nudef}) by (\ref{genfin}). Laplace’s method is based on the exponential character of the integrand in the large $\Pr$ limit. As $\Pr\to\infty$, the magnitude of the integrand becomes concentrated near the location $u = 1$, which corresponds to the maximum of the exponential argument. This justifies the use of the Taylor series about $u = 1$ in the method (see (\ref{phikdef})) and in subsequent simplifications. We recall here that $u = \exp(-C\eta/2)$ (see text just above  (\ref{genfin}), or equivalently Eqs. (\ref{atildeexpmain}) and (\ref{gdef})), and thus the limit as $u \to 1$ corresponds to that of $\eta \to 0$ in (\ref{mom}).  Thus, we hypothesize---and will go on to demonstrate---that the full Taylor series expansion of the solution in $\eta$ about $\eta = 0$ may be used directly to obtain the same large Pr expansion (\ref{gexp}) and coefficients in Appendix B.

The power series solution of the Sakiadis problem (\ref{mom}) about $\eta=0$ is well established, and in fact, the recursion for the coefficients was first given by~\citet{Blasius} for flow over a stationary plate.  That solution was modified slightly by~\citet{Sakiadis1961b} to satisfy the different boundary conditions in the Sakiadis flow, and is given in Appendix C by (\ref{flatser}).  The solution has a finite radius of convergence, $R$---the solution converges only when $|\eta| < R$, where $R \approx 4.0722$, and is divergent for larger $\eta$ values.  Hence, the series solution cannot bridge the semi-infinite domain in the Sakiadis problem (\ref{mom}), and is thus deficient.  It is for precisely this reason that~\citet{sak} developed the alternative convergent expansion given by (\ref{welpthismyman}).  Nevertheless, in Appendix C, we show that, in fact, the divergent infinite series expansion (\ref{flatser}) may be used to extract the \textit{full} asymptotic expansion (\ref{gexp}) with identical coefficients shown in Appendix A.  The methodology using the divergent power series solution takes less steps (compare work in Appendices B and C) and satisfies a slightly simpler recurrence relation than that using the exact convergent solution.  

Thus, we call out this interesting element, namely that it is not necessary to have a convergent solution for the full flow field to obtain the full $\Pr \to \infty$ asymptotic behavior. Only the flow field near the wall needs to be represented accurately, and the series divergence away from the wall does not impact Laplace’s method---as the method focuses on the dominant behavior of the solution as $\Pr \to \infty$ near the wall.  The fact that the divergent series expansion may be used in the $\Pr \to \infty$ limit is a physically appealing result, since the characteristic thickness of the thermal boundary layer is small compared with that of the hydrodynamic boundary layer for large $\Pr$. This is precisely as discussed in the approximate analysis shown in Section (3.1).  In Section 2, recall that we mentioned it was surprising that the asymptotic constant $C$, where $f\sim C$ as $\eta \to \infty$, did not appear in the $\Pr \to \infty$ asymptotic behavior when the convergent expansion (\ref{welpthismyman}) was employed.  We see now why that is the case---the use of the divergent expansion makes clear that it is only the behavior of the flow solution near the wall that matters to the asymptotic solution, and thus, only the shear parameter $\kappa$ appears in the asymptotic coefficients (see Appendix B).
	
 Finally, we comment on the use of polynomial approximations to the velocity field used to predict convective heat transfer coefficients in boundary layer flows. In order to reproduce the correct Laplace's method expansion as $\Pr \to \infty$, it is necessary that a power series be generated as an expansion about the wall ($\eta=0$) in accordance with the differential equation itself, as the power series solution to (1). It is commonplace for zero slope, curvature, etc. to be applied at the edge of the boundary layer when using a Polhausen approximation to predict wall shear stresses; see, for example~\citet{Schlichting} (page 206).  However, one cannot utilize such an approach when coupling the polynomial approximation with a thermal boundary layer problem if one wants to recover the correct asymptotic behavior as $\Pr \to \infty$.  It is only if the power series is generated algorithmically as described above,  using information at the wall for higher derivatives, that the correct polynomials can be generated with a finite number of terms.

\begin{figure}
\centering
\includegraphics[width=\textwidth]{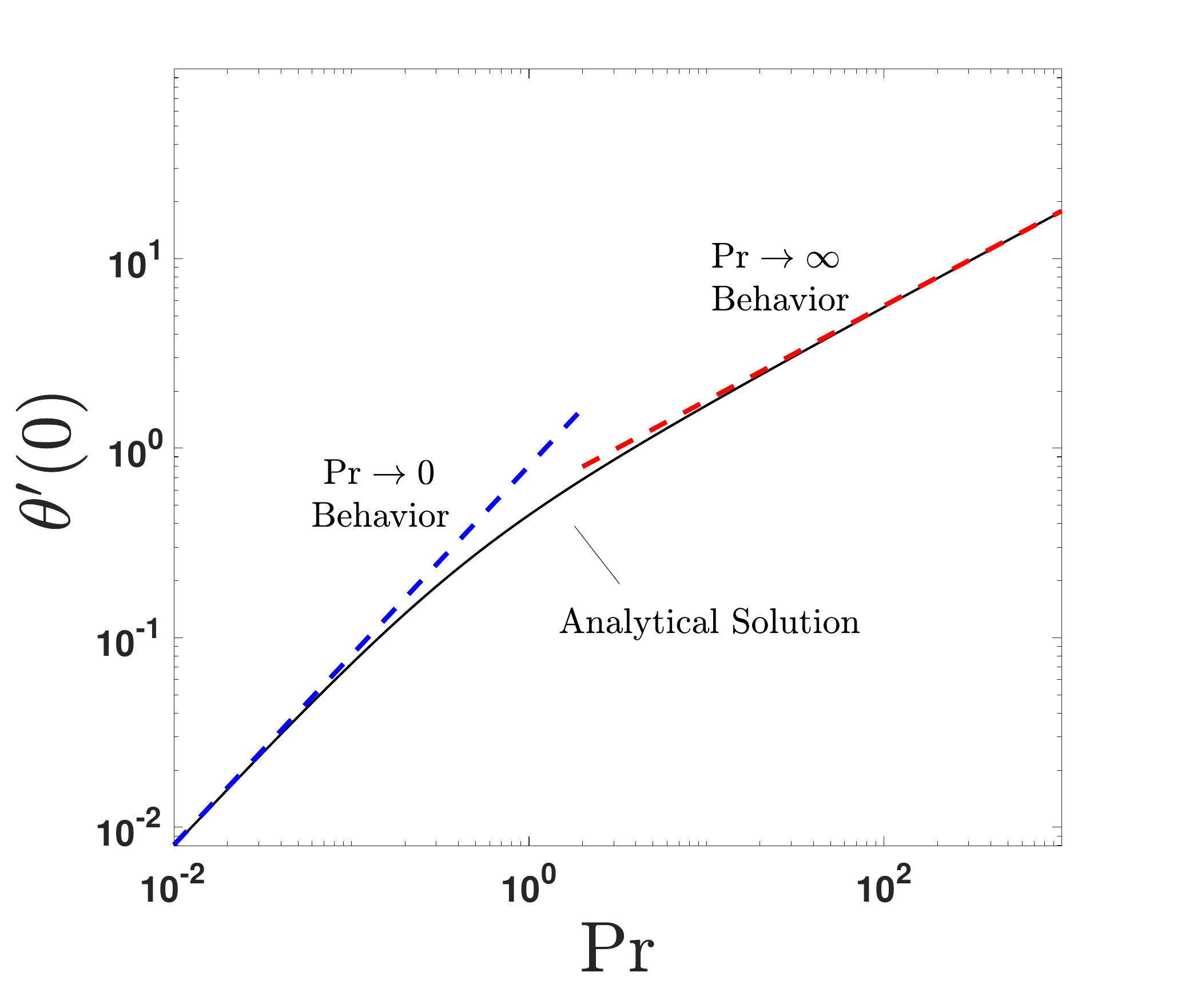}
\caption{Plot showing the effect of $\Pr$ on $\tdz$ from the analytical solution (\ref{thebigsum}). The asymptotic behavior $\tdz \sim C \Pr/2 ~ (\Pr \to 0)$ and $\tdz \sim \pi^{-1/2} \Pr^{1/2} ~ (\Pr \to \infty)$ are shown as dashed lines on the plot for comparison.}
\label{thetaPrLog}   
\end{figure}

\begin{figure}
\centering
\includegraphics[width=\textwidth]{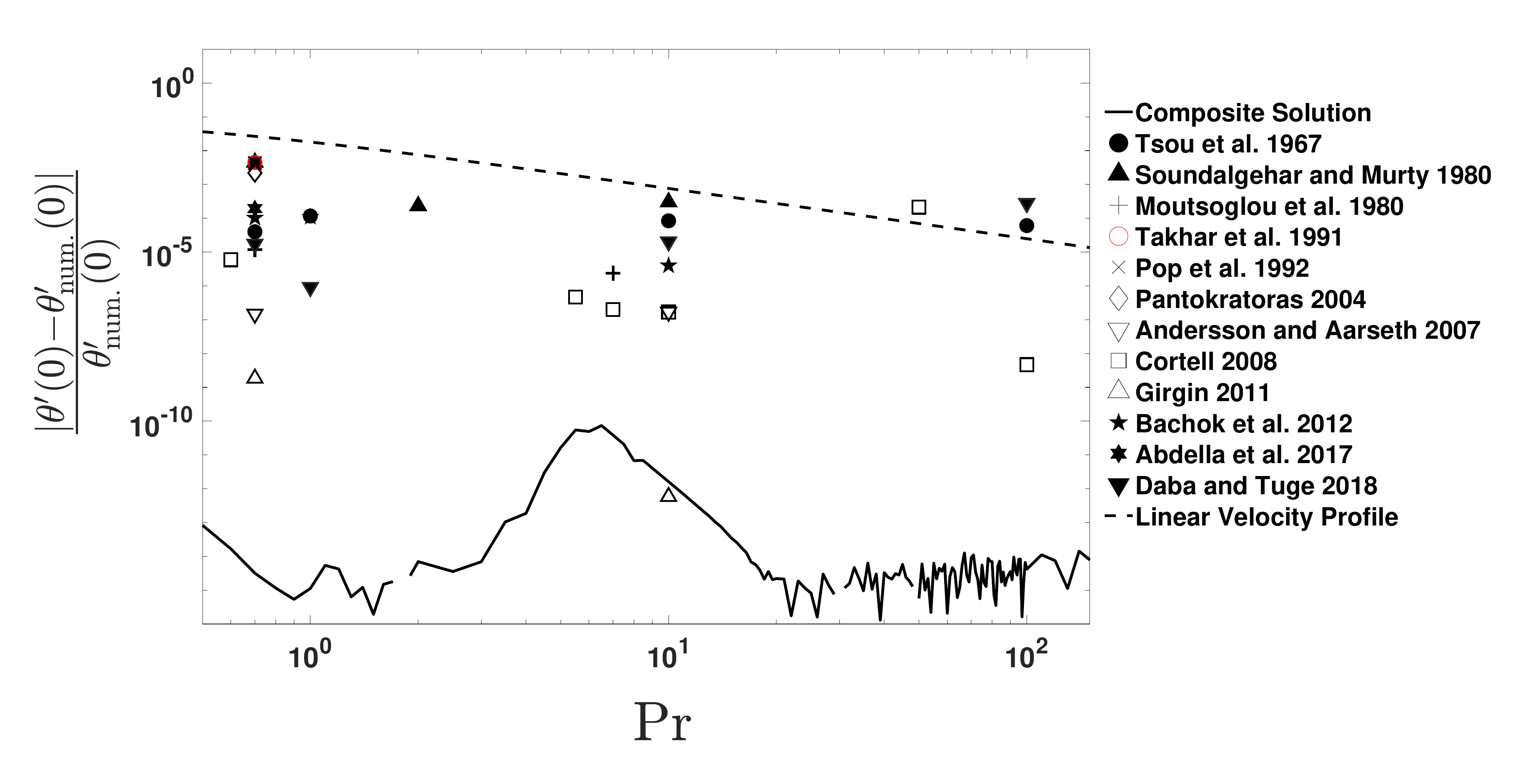}
\caption{Relative error in $\tdz$ between the present model (\ref{thebigsum}) and numerical solution of (\ref{keyint}). The error in numerically predicted $\tdz$ from prior literature \cite{Tsou1967,Sound1980,Mout1980,Takhar1991,Pop1992,Pant2004,And2007,Cortell2008,Girgin2011,Bachok2012,Abd2017, Daba2018} is presented for comparison, as well as that of the linear velocity profile, which is a common first approximation for this type of problem. Note that the values given in previous work are restricted to discrete values of $\Pr$, contrasting with Figure \ref{thetaPrLog}.}
\label{ErThetaDPlot}   
\end{figure}

Figure \ref{thetaPrLog} shows the effect of Prandtl number on $\tdz$ calculated using the analytical result (\ref{thebigsum}) as well as the asymptotic behaviors at large and small $\Pr$. (\ref{dexp}) indicates that the behavior of the temperature gradient, $\tdz$, scales as $\tdz\sim \pi^{-1/2}\Pr^{1/2}$ as $\Pr \to \infty$; similarly, as $\Pr \to 0$, $\tdz \sim C \Pr /2$ from (\ref{sergen}). When plotted on Figure \ref{thetaPrLog}, the prior literature results from Table 1---that are numerical and valid for discrete values of $\Pr$---are indistinguishable from the curve generated from (\ref{thebigsum}). Figure \ref{ErThetaDPlot} provides an error analysis of our analytical model compared with machine precision numerical results, and here superimposes associated errors from prior studies from Table 1. The figure indicates that the present analytical model is accurate to $O(10^{-10})$ when compared with the numerical results for all values of Prandtl number. We have also superimposed the result (\ref{thetaD1111}), obtained using a linear velocity approximation, on Figure \ref{ErThetaDPlot}.  Since such a low-order polynomial is often utilized in boundary layer estimates, the benefit of using more terms in the asymptotic expansion for large $\Pr$ is observed (recall that the $\Pr\to\infty$ asymptotic expansion is used in the analytical result (\ref{thebigsum}) for $\Pr\geq 6$). 

\subsection{Importance of the wall shear stress}
\label{Results}
The presented exact and approximate models in Sections~\ref{Exact} and \ref{ApproxMod} are strictly dependent on the dimensionless wall shear stress parameter, $\kappa$ in the large $\Pr$ regime, and have more complicated physics in the small $\Pr$ regime. From our study, it is apparent why accuracy in the $\kappa$ affects that of $\tdz$ as evidenced by the work cited in Table 1. Inspection of the coefficients $\gamma_n$  in (\ref{gexp}) and written explicitly in Appendix B show that they are solely a function of $\kappa$ as $\Pr\to\infty$.  This again reinforces that the behavior of the flow field near the wall dominates the asymptotic behavior in this limit. Among previous studies, the results of  $\tdz$ calculated by~\citet{Girgin2011} at different $\Pr$ are the closest to those predicted by the present analytical model (with an error as low as $10^{-9}$) as indicated in Figure \ref{ErThetaDPlot}. This is connected to the agreement between the value of $\kappa$ given by~\citet{Girgin2011} and that used in our model given explicitly by~\citet{exactPreprint}. On the other hand, the results calculated by other studies~\cite{Pop1992,Takhar1991,Pant2004}, which have large error in $\kappa$, have large error in $\tdz$ as well (see Figure \ref{ErThetaDPlot}). The importance of having an accurate value of $\kappa$ is clearly indicated from its explicit appearance in the asymptotic form (see asymptotic constants in appendix B).




\section{Conclusions}
\label{Conc}  
We have utilized a known exact solution for the Sakiadis boundary layer to develop a similarly exact solution for the associated thermal boundary layer. Although exact, we find that the solution for the temperature gradient at the wall, needed to extract the convective heat transfer coefficient, is computationally unstable at large $\Pr$.  To compensate, we developed a large $\Pr$ expansion for the temperature gradient using Laplace's method.  By combining the exact and asymptotic solutions, an analytical expression for the temperature gradient is obtained that is accurate to $O(10^{-10})$  for all $\Pr$ when compared with machine-precision numerical results.  The large $\Pr$ asymptotic analysis shows that a divergent power series solution for the fluid flow---that is itself expanded about the location of the moving surface---may be used instead of the convergent power series solution to obtain all higher order coefficients in the asymptotic expansion.  We show that this result is connected to the physics of large Prandtl number flows where the thickness of the hydrodynamic boundary layer is much larger than that of the thermal boundary layer. The latter approach must be used when trying to estimate the wall shear stress using the Polhausen approximation, but is not the appropriate approximation to use for the heat transfer asymptotics. Additionally, to remain consistent with the $\Pr \to \infty$ asymptotic behavior, we have shown that any higher-order constraints on an assumed polynomial velocity field used to approximate heat transfer must be consistent with the boundary-layer equations evaluated at the wall, as opposed to imposing higher-order constraints at the edge of the boundary layer itself.  We also demonstrate that a divergent infinite power series expanded about the wall location may be used in place of an exact convergent series expansion to examine the large $\Pr$ limit via Laplace's method. By the well-known analogy between heat and mass transfer in boundary layer flows, the results of our study may also be used to describe the convective mass transport of a dilute solute into the fluid from the wall.  The present model is highly accurate and attractive for ease of use in engineering applications.

\appendix 

\section{Computation of the Asymptotic Coefficients}
\label{AppA}

In this appendix, we provide details of Laplace's method used to evaluate the integral $I$ in (\ref{intlapform}) as $\Pr \to \infty$, which underpins the expressions (\ref{gexp}) and (\ref{dexp}) in the text. This process proceeds in two steps. In Appendix~\ref{AsymLap}, a recursive formula is derived for all coefficients of the large $\Pr$ asymptotic expansion of the integral $I$. In Appendix~\ref{Alter}, we express the recursive formula in terms of the wall shear stress extracted from the Sakiadis boundary layer flow. 

\subsection{Asymptotic Expansion of $I$ via Laplace's method}
\label{AsymLap}

\begin{figure}
    \centering
    \includegraphics[width=\textwidth]{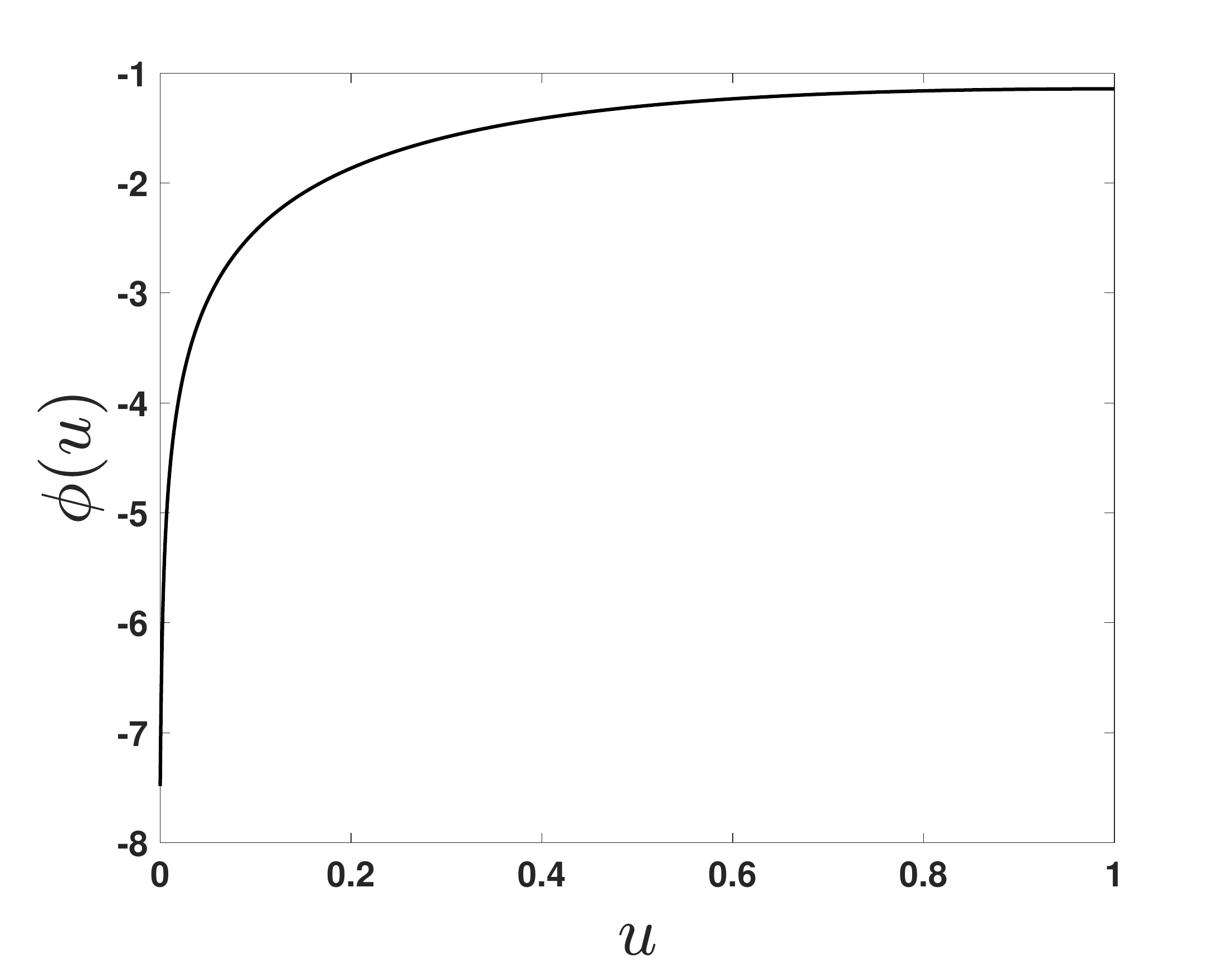}
    \caption{Behavior of the Function $\phi$ in the domain of integration.}
    \label{fig:phiplot}
\end{figure}

The form of the integral (\ref{intlapform}) suggests that Laplace's method may be used to extract its large $\Pr$ behavior. As observed in Figure \ref{fig:phiplot} and from explicit calculations in Appendix~\ref{Alter}, we see that $\phi(u)$ has a maximum at $u=1$ for which $\frac{\text{d}\phi}{\text{d}u} = 0$. As a consequence of the exponential term in the integrand, the integral in (\ref{intlapform}) is governed its behavior near $u=1$ for large $\Pr$. To facilitate asymptotic expansion, we Taylor expand the integrand about $u=1$ as

\begin{subequations}
\begin{align}
    \phi(u) &= \sum_{n=0}^\infty \phi_n (u-1)^n, \label{phikdef} \\ 
    \frac{1}{u} &= \sum_{n=0}^\infty (-1)^n (u-1)^n.
\end{align}
\end{subequations} The Taylor coefficients $\phi_n$ of $\phi(u)$ about $u=1$ can be computed directly by differentiating the expression (\ref{phidef}) term-by-term to yield \begin{align}\begin{split} \phi_k & = \frac{1}{k!} \left.\left[\frac{\text{d}^k\phi}{\text{d}u^k}\right]\right|_{u=1} = \frac{1}{k!} \left.\left[\frac{\text{d}^k}{\text{d}u^k}\left(\log(u) + \sum_{n=1}^\infty \frac{\tilde{a}_n}{nC} u^n\right)\right]\right|_{u=1} \\ & = \frac{(-1)^{k+1}}{k} + \sum_{n=0}^\infty \binom{n+k}{k} \frac{\tilde{a}_{n+k}}{(n+k)C}. \label{phimyman}\end{split}\end{align} In Appendix~\ref{Alter}, we provide an alternative recursive method to calculate these coefficients that eliminates the need to evaluate the infinite series in (\ref{phimyman}). In (\ref{phidef}), note that $\phi_0 = \phi(1), \phi_1 = 0, \phi_2 < 0$ (this is consistent with the curve shape given shown in Fig~\ref{fig:phiplot}). As the dominant contribution of the integral in (\ref{jjettasiguessimouttajokes}) occurs near $u=1$, we examine the integral in the neighborhood of $u=1$ as

\begin{subequations}
\begin{align} \nonumber I &\sim \int_{1-\varepsilon}^1 \left(\sum_{n=0}^\infty (-1)^n (u-1)^n\right)\exp \biggl(\Pr\phi_0+\Pr\phi_2(u-1)^2  \\ &~~~~~~~~~~~~~~~~~~~~~~~~~~+ \Pr\sum_{k=3}^\infty \phi_k(u-1)^k\biggr) \ \text{d}u \\ \nonumber &\sim e^{\phi_0\Pr} \sum_{n=0}^\infty (-1)^n \int_{1-\varepsilon}^1 (u-1)^n e^{\Pr\phi_2(u-1)^2}  \\ &~~~~~~~~~~~~~~~~~~~~~~~~~~ \times ~~\exp\left(\Pr\sum_{k=3}^\infty \phi_k(u-1)^k \right) \ \text{d}u \\ &\sim e^{\phi_0 \Pr} \sum_{n=0}^\infty (-1)^n\int_{-\varepsilon}^0 v^n e^{\Pr\phi_2 v^2} \exp\left(\Pr\sum_{k=3}^\infty \phi_k v^k\right) \ \text{d}v. \end{align} \label{lapstepone}
\end{subequations}

In (\ref{lapstepone}), note that $\varepsilon > 0$ is a small value that justifies the use of the Taylor series to represent the integrand in the large $\Pr$ limit.  As a result, the last exponential in (\ref{lapstepone}) may be expanded as a Taylor about $v=0$, giving the expressions

\begin{subequations}
\begin{align} I &\sim e^{\phi_0 \Pr} \sum_{n=0}^\infty (-1)^n \int_{-\varepsilon}^0 v^n e^{\phi_2\Pr v^2} \sum_{k=0}^\infty \frac{\Pr^k}{k!}\left\{\sum_{m=3}^\infty \phi_{m} v^m\right\}^k  \ \text{d}v \label{firstphiexp} \\ &\sim e^{\phi_0 \Pr} \sum_{n=0}^\infty (-1)^n \int_{-\varepsilon}^0 v^n e^{\phi_2\Pr v^2} \sum_{k=0}^\infty \frac{\Pr^k}{k!}\left\{\sum_{m=3k}^\infty \varphi_{k, m} v^m\right\} \ \text{d}v \label{secondphiexp} \\ &\sim e^{\phi_0\Pr} \sum_{n=0}^\infty \sum_{k=0}^\infty \sum_{m=3k}^\infty \frac{(-1)^n}{k!} \varphi_{k, m} \Pr^k \int_{-\varepsilon}^0 v^{n+m} e^{\phi_2 \Pr v^2} \ \text{d}v \label{lapsteptwoc} \end{align} where \begin{equation}
    \left[\sum_{m=3}^\infty \phi_{m} v^m\right]^k = \sum_{m=3k}^\infty \varphi_{k, m} v^m.  \label{varphidef}\end{equation} \label{ringwald}
\end{subequations} Note that the expansion (\ref{varphidef}) is used in moving from (\ref{firstphiexp}) to (\ref{secondphiexp}), and simplifies the individual integral terms in (\ref{lapsteptwoc}) to be eventually identified as Gamma functions (see (\ref{gammaint}) below). Using standard series techniques, all $\varphi_{k, m}$ can be computed recursively using the Cauchy product of series as \begin{subequations}
\begin{align} \varphi_{0, 0} &= 1, \hphantom{\sum\sum\sum} \varphi_{0, m} = 0, m \geq 1, \\ \varphi_{k, m} &= \sum_{j=3}^m \phi_j \varphi_{k-1, m-j}, k \geq 1, m \geq 0. \end{align}
\end{subequations}

 Next, Laplace's method allows us to extend the limit of integration to $\varepsilon = \infty$, as it incurs subdominant contributions to the $\Pr \to \infty$ asymptotic behavior as can be established with integration by parts. The integral in (\ref{lapsteptwoc}) is subsequently evaluated as \begin{multline} \int_{-\varepsilon}^0 v^{n+m} e^{\phi_2 \Pr v^2} \ \text{d}v \sim \int_{-\infty}^0 v^{n+m} e^{\phi_2 \Pr v^2} \ \text{d}v  \\ = \frac{(-1)^{n+m}}{2}\left(-\phi_2\Pr \right)^{\frac{-n-m-1}{2}} \Gamma\left(\frac{n+m+1}{2}\right), \label{gammaint} \end{multline} where $\Gamma(z)$ is the Gamma function. Solving (\ref{gammaint}) with (\ref{ringwald}) we obtain \begin{equation} I \sim e^{\phi_0\Pr} \sum_{n=0}^\infty \sum_{k=0}^\infty \sum_{m=3k}^\infty \frac{\displaystyle (-1)^m \varphi_{k, m} \Gamma\left(\frac{n+m+1}{2}\right)}{\displaystyle 2(-\phi_2)^{(n+m+1)/2}k!} \Pr^{(2k-n-m-1)/2}. \label{trip_sum_to_reorder} \end{equation}

The final step is to collect terms of the same order in $\Pr$. To do this, we note that the summation indices $(n, k, m)$ form precisely the integer points in $\mathbb{Z}^3$ satisfying the trio of inequalities $n \geq 0, k \geq 0, m \geq 3k$. The idea is to change basis on $\mathbb{Z}^3$ so that the main term in the exponent of $\Pr^{-1}$ in the expansion (\ref{trip_sum_to_reorder}), $n+m-2k$, becomes a new basis element. This could be done in many ways, but for definiteness we note one with the change of basis $$ (n, k, m) \mapsto (n, k, n+m-2k).$$ This works because it makes the main term a basis element, but it maps integer points to integer points and vice versa because the inverse of this change of basis matrix $$ \begin{pmatrix} 1 & 0 & 0 \\ 0 & 0 & 1 \\ 1 & 1 & -2 \end{pmatrix}^{-1} = \begin{pmatrix} 1 & 0 & 0 \\ -1 & 2 & 1 \\ 0 & 1 & 0 \end{pmatrix} $$ has integer coefficients. From this, one gets the result. \begin{subequations} \begin{align} \gamma_\ell &= \sum_{n=0}^{\ell-1} \sum_{k=0}^{\ell-n-1} \frac{\displaystyle (-1)^{\ell+2k-n-1} \varphi_{k, \ell+2k-n-1} \Gamma\left(\frac{\ell+2k}{2}\right)}{\displaystyle 2(-\phi_2)^{(\ell + 2k)/2}k!}. \label{gammarecforreal} \\ \varphi_{0, 0} &= 1, \hphantom{\sum\sum\sum} \varphi_{0, m} = 0, m \geq 1, \\ \varphi_{k, m} &= \sum_{j=3}^m \phi_j \varphi_{k-1, m-j}, k \geq 1, m \geq 0. \end{align} \label{gammarecforrealreal} \end{subequations} (\ref{gammarecforrealreal}) provides the coefficients for the asymptotic expansion (\ref{gexp}) to all orders in $\Pr$. 


\subsection{Alternative Evaluation of the Taylor Coefficients of $\phi$ in (\ref{phikdef})}
\label{Alter}

As noted previously, the recursion from the last Section for the Laplace expansion of the integral used in determining $\tdz$ requires the Taylor coefficients of the function $\phi(u)$ at $u=1$, denoted as $\phi_k$ in (\ref{phikdef}). (\ref{phimyman}) expresses a typical determination of the Taylor coefficients by differentiation and subsequent evaluation at $u=1$. Here we eliminate the need for the infinite sums in the computation of the coefficients provided in (\ref{phimyman}). 

For simplicity, we first define a function $q(u)$ by rearranging (\ref{intlapform}) as \begin{equation} q(u) = \sum_{n=1}^\infty \frac{\tilde{a}_n}{nC}u^n = \phi(u) - \ln(u). \label{qdef} \end{equation} The $k$-th Taylor coefficient of $\ln(u)$ about $u=1$ is well-known to be $(-1)^{k+1}/k$, so to compute the coefficients $\phi_k$ it suffices to compute the Taylor coefficients of $q(u)$ about $u=1$. This, in turn, is done by establishing a relationship between $q$, and the dimensionless stream function $f$ given by (\ref{welpthismyman}). Substituting the relationship $u = \exp(-C\eta/2)$ that was used in the derivation of (\ref{keyint}) into the expression for $f$ in (\ref{welpthismyman}) we write \begin{equation} f(\eta) = \sum_{n=0}^\infty \tilde{a}_n e^{-Cn\eta/2} = g(e^{-C\eta/2}) \label{atildeexpmain}\end{equation} so that \begin{equation}  g(u) = f\left(-\frac{2}{C}\ln(u)\right) = \sum_{n=0}^\infty \tilde{a}_n u^n. \label{gdef} \end{equation} The coefficients $\tilde{a}_n$ are provided in (\ref{welpthismyman}) above. Note that these coefficients are obtained by~\citet{sak} by transforming the third-order Sakiadis ODE in $\eta$ to $u$ and subsequently expanding about $u=0$ ($\eta=\infty$). Note that the coefficients $\phi_n$ in (\ref{phikdef}), however, correspond to a Taylor expansion about $u=1$ ($\eta = 0$). 

A key observation is that the differentiation of $q$ in (\ref{qdef}) term-by-term eliminates the $1/n$ factor in each term in the infinite sum, so $q'$ corresponds to $g/C$, except that the coefficients are shifted by one, which we can fix by multiplying by $u$ and adding in the first coefficient $\tilde{a}_0$. That is, by examining the Taylor series at $u=0$ we note that $Cuq'(u) + \tilde{a}_0$ has the same Taylor expansion as $g$, and hence $Cuq'(u)+\tilde{a}_0 = g(u)$. Since $\tilde{a}_0 = C$ this differential equation for $q$ can be integrated to yield \begin{equation} q(u) = q(1) + \int_1^u \frac{g(\tau) - C}{C\tau} \ \text{d}\tau. \label{gqrelation} \end{equation} At this point we can evaluate the integral (\ref{gqrelation}) using the Taylor expansion of $g(u)$ about $u=1$.  This is also derived by Naghshineh et al. \cite{sak} as \begin{subequations}\begin{equation} g(u) = \sum_{n=0}^\infty \hat{a}_n (u-1)^n\end{equation} where
    \begin{align} \begin{split} \hat{a}_{n+3} &= -\frac{2n+3}{n+3}\hat{a}_{n+2} - \frac{(n+1)^2}{(n+2)(n+3)} \hat{a}_{n+1} \\ & +\frac{1}{C(n+1)(n+2)(n+3)}  \biggl\{\sum_{k=0}^{n-2}(k+1)(k+2)\hat{a}_{k+2}\hat{a}_{n-k-1}  \\ & ~~~~~~~~~~~~~~~+ \sum_{k=0}^{n-1}(k+1)\left[(k+2)\hat{a}_{k+2} + \hat{a}_{k+1}\right]\hat{a}_{n-k} \biggr\}, \\ \hat{a}_0 &= 0, \hphantom{\sum} \hat{a}_1 = -\frac{2}{C}, \hphantom{\sum} \hat{a}_2 = \frac{2\kappa}{C^2} + \frac{1}{C}. \label{ahatrec} \end{split}\end{align}
\end{subequations}

We can thus use standard series manipulations of the integrand to derive the Taylor coefficients of $q(u)$ about $u=1$; this provides us with a method to compute the desired coefficients $\phi_k$ in (\ref{phidef}) as follows. For the first term, $q(1) = \phi(1) = \phi_0$ in (\ref{phidef}); we cannot compute this value without summing an infinite series (although, as stated before, it converges to double precision in $35$ terms). However, its value is not needed as it does not appear in (\ref{dexp}) after cancellation of the exponential that appears in the reciprocal form (\ref{gexp}). All of the higher order terms, though, can be computed by Taylor expanding $g(\tau)-C$ about $\tau=1$ in the integrand, expanding the $\tau^{-1}$ as a geometric series in $\tau-1$, computing the Cauchy product of these series (which simply gives the Taylor expansion of the product of two functions with known expansions), and integrating term-by-term. In particular, if we first write \begin{equation} \frac{g(\tau) - \tau}{C} = \sum_{n=0}^\infty \hat{\hat{a}} (\tau-1)^n, \end{equation} we then have $\hat{\hat{a}}_0 = -1$ and $\hat{\hat{a}}_n = \hat{a}_n/C$. If we further write \begin{equation} \frac{g(\tau)-C}{C\tau} = \frac{g(\tau)-C}{C(1+(\tau-1))} = \sum_{n=0}^\infty \tilde{\alpha}_n (\tau-1)^n, \end{equation} we can simply compute the series product of the sum of the $\hat{\hat{a}}_n$ and the geometric series from the denominator as \begin{equation} \tilde{\alpha}_n = \sum_{k=0}^n \hat{\hat{a}}_k (-1)^{n-k}. \end{equation} We subsequently integrate this result term-by-term to obtain \begin{equation} q(u) = q(1) + \int_1^{u} \sum_{n=0}^\infty \tilde{\alpha}_n(\tau-1)^n \ \text{d}\tau = q(1) + \sum_{n=0}^\infty \frac{\tilde{\alpha}_n}{n+1} (u-1)^{n+1}. \label{alphahatuse} \end{equation} Rearranging (\ref{qdef}), we have \begin{equation} \phi(u) = q(u) + \ln(u). \label{phiredef} \end{equation} To extract the Taylor coefficients $\phi_n$ according to (\ref{phikdef}), then, we require the Taylor expansion of $\ln(u)$ as \begin{equation} \ln(u) = \sum_{n=1}^\infty \frac{(-1)^{n+1}}{n}(u-1)^n. \label{mercator} \end{equation} Inserting Eqs.~(\ref{alphahatuse}) and (\ref{mercator}) into (\ref{phiredef}), and comparing with the form (\ref{phikdef}) term by term, we obtain the desired expression \begin{equation} \phi_n = \frac{1}{n}\left(\tilde{\alpha}_{n-1} + (-1)^{n+1}\right), n\geq 1.  \end{equation}

As a matter of summary and for convenience in referencing final results, we compile necessary components of $\phi_n$ as follows \begin{subequations} \begin{align}
    \phi_0 &= \phi(1) = \sum_{n=1}^\infty \frac{\tilde{a}_n}{nC}, \\
    \phi_n &= \frac{1}{n}\left(\tilde{\alpha}_{n-1} + (-1)^{n+1}\right), n\geq 1,  \\ \tilde{\alpha}_n &= \sum_{k=0}^n \hat{\hat{a}}_k (-1)^{n-k}, n \geq 0, \\ \hat{\hat{a}}_0 &= -1 \hphantom{\sum\sum\sum} \hat{\hat{a}}_n = \hat{a}_n/C, n \geq 1 \\   
    \begin{split} \hat{a}_{n+3} &= -\frac{2n+3}{n+3}\hat{a}_{n+2} - \frac{(n+1)^2}{(n+2)(n+3)} \hat{a}_{n+1} \\ & +\frac{1}{C(n+1)(n+2)(n+3)}  \biggl\{\sum_{k=0}^{n-2}(k+1)(k+2)\hat{a}_{k+2}\hat{a}_{n-k-1}  \\ & ~~~~~~~~~~~~~~~+ \sum_{k=0}^{n-1}(k+1)\left[(k+2)\hat{a}_{k+2} + \hat{a}_{k+1}\right]\hat{a}_{n-k} \biggr\}, \\ \hat{a}_0 &= 0, \hphantom{\sum} \hat{a}_1 = -\frac{2}{C}, \hphantom{\sum} \hat{a}_2 = \frac{2\kappa}{C^2} + \frac{1}{C}. \label{ahatrec} \end{split}
\end{align} \label{phifin} \end{subequations}  At this point, with the $\phi_n$ coefficients determined in (\ref{phifin}), the Laplace method coefficients $\gamma_n$ and $\delta_n$ in Eqs.~(\ref{gexp}) and (\ref{dexp}) can be easily determined recursively. Note that the wall shear $\kappa$ and constant $C$ are introduced into the function $g(u)$ upstream in the calculation at (\ref{ahatrec}), and these constants are embedded in the final recursion (\ref{phifin}). Because there are relatively few terms before the optimal truncation, a computer algebra system was used to determine non-recursive expressions for the coefficients $\gamma_n$ used in (\ref{dexp}) and given in Appendix~\ref{GamTab}. It is notable that the constant $C$ does not appear in the final expressions for the coefficients $\gamma_n$. We have verified that  the constants in Appendix~\ref{GamTab} do agree with the direct implementation of (\ref{gammarecforreal}) and with the $\phi_n$ determined non-recursively according to (\ref{phimyman}). This self-consistency is essential as it is not obvious by inspection of the recursive formula (\ref{gammarecforrealreal}) that the $\gamma_n$ should be independent of $C$. This is also confirmed by the use of an alternative implementation of Laplace's method described in Appendix~\ref{C}.

\section{Table of $\gamma_n$}
\label{GamTab}

\begin{widetext}

\begin{align*}
    \gamma_1 &= \sqrt{\pi} \\[4mm]
    \gamma_2 &= -\frac{2}{3}\kappa \\[4mm]
    \gamma_3 &= \sqrt{\pi}\frac{5}{12} \kappa^2 \\[4mm]
    \gamma_4 &= -\frac{1}{135}\left[ 160\kappa^3-18\kappa \right] \\[4mm]
    \gamma_5 &= \frac{\sqrt{\pi}}{288}\left[385\kappa^4-72\kappa^2 \right] \\[4mm]
    \gamma_6 &= -\frac{1}{2835}\left[15680\kappa^5-4032\kappa^3+162\kappa\right] \\[4mm]
    \gamma_7 &= \frac{\sqrt{\pi}}{51840}\left[425425\kappa^6 - 138600\kappa^4 + 10152\kappa^2\right] \\[4mm]
    \gamma_8 &= - \frac{1}{8505}\left[358400\kappa^7 - 141120\kappa^5 + 14975\kappa^3 - 270\right] 
    \\[4mm]
    \gamma_9 &= \frac{\sqrt{\pi}}{2488320}\left[185910725\kappa^8-85765680\kappa^6+11927520\kappa^4-445824\kappa^2\right] \\[4mm]
    \gamma_{10} &= -\frac{1}{1804275}\left[805376000\kappa^9-425779200\kappa^7-73275840\kappa^5+4230144\kappa^3-36450\kappa\right] \\[4mm]
    \gamma_{11} &= \frac{\sqrt{\pi}}{209018880}[188699385875\kappa^{10}-112438806480\kappa^8+23070967920\kappa^6-1814206464\kappa^4 \\ & 
\hphantom{\sum\sum\sum\sum\sum\sum\sum\sum\sum\sum\sum\sum\sum\sum\sum\sum\sum\sum\sum} +38257920\kappa] \\[4mm]
    \gamma_{12} &= -\frac{1}{492567075}[2984869888000\kappa^{11} - 1978808832000\kappa^9 + 471593041920\kappa^7 - 47103166656\kappa^5 \\ & \hphantom{\sum\sum\sum\sum\sum\sum\sum\sum\sum\sum\sum\sum\sum\sum\sum\sum} +1650572640\kappa^3 - 6889050\kappa]  \\[4mm]
    \gamma_{13} &= \frac{\sqrt{\pi}}{75246796800}[1023694168371875\kappa^{12} - 747249568065000\kappa^{10}  \\ & \hphantom{\sum\sum} +202871732263200\kappa^8 - 24617993662080\kappa^6 + 1225752932736\kappa^4 - 15340492800\kappa^2] \\[4mm]
    \gamma_{14} &= -\frac{1}{1477701225}[148325072896000\kappa^{13} - 118200847564800\kappa^{11} + 36014320742400\kappa^9 \\ & \hphantom{\sum\sum\sum\sum\sum}- 5148878192640\kappa^7 + 334802429568\kappa^5 - 7547494464\kappa^3 + 15155910\kappa] \\[4mm]
    \gamma_{15} &=\frac{\sqrt{\pi}}{902961561600}[221849150488590625\kappa^{14} - 191635548319215000\kappa^{12} + 64754512569747000\kappa^{10} \\ & \hphantom{\sum} - 10664946375632640\kappa^8 + 859343139639936\kappa^6 - 28793424927744\kappa^4 + 220085683200\kappa^2]
\end{align*}

\end{widetext}

%

We note that $\kappa = f''(0)$ is the wall-shear, which has numerical value $\kappa = -0.443748313368861...$, which was expressed in closed form by \cite{exactPreprint} and is given in (\ref{exKappa}) of the main text.

\section{Alternative Approach: Large $\Pr$ asymptotic expansion via divergent Sakiadis Power Series Solution}
\label{C}

In Appendix~\ref{AsymLap}, Laplace's method was used to determine the large $\Pr$ behavior of the integral (\ref{intlapform}); this integral is embedded in the desired Nusselt number expression (\ref{nudef}) by (\ref{keyint}).  The purpose of this appendix is to show that one may utilize the Taylor series expansion in $\eta$ about $\eta=0$ directly to obtain the same large $\Pr$ expansion (\ref{gexp}) and coefficients in Appendix~\ref{GamTab}.

To proceed, we make use of the well-established power series solution to (\ref{firstbruh}), with a recursion for the coefficients given first by~\citet{Blasius} for flow over a stationary plate, but here written for the Sakiadis flow as  \begin{subequations}
\begin{align} 
    f(\eta) &= \sum_{n=0}^\infty a_n \eta^n, \label{flatserdef}\\ a_{n+3} &= -\frac{\displaystyle\sum_{k=0}^n(k+1)(k+2)a_{k+1}a_{n-k}}{2(n+1)(n+2)(n+3)}, \hphantom{\sum} n \geq 0 \\   a_0 = 0, \hphantom{\sum}  a_1 &= 1, \hphantom{\sum} a_2 = \kappa/2  . 
\end{align} \label{flatser}
\end{subequations} As noted by~\citet{sak}, the series (\ref{flatser}) is convergent in the range $\eta<R$, where $R\approx4.0722$. Since the series (\ref{flatser}) for $f$ diverges for $\eta > R$, the series solution is deficient in that it cannot satisfy the boundary condition $f' \to 0$ as $\eta \to \infty$ in (\ref{firstbruh}).  In fact, this was the motivation for the gauge transformation that yields the convergent solution (\ref{welpthismyman}) (see \cite{sak}).  Nevertheless, the series (\ref{flatser}) \textit{is} sufficient to extract the large $\Pr$ behavior in accordance with Laplace's method as follows.  

For Large $\Pr$, the integral is dominated by its integrand's behavior near $\eta=0$; thus, we first replace the upper bound in (\ref{genfin}) with a small parameter $\varepsilon > 0$ which justifies the use of the expansion (\ref{flatserdef}) in the evaluation of the integral.  

\begin{equation}
    \tdz^{-1} \sim \int_0^\varepsilon \exp\left(-\frac{1}{2}\Pr\sum_{n=1}^\infty \frac{a_n}{n+1}\eta^{n+1}\right) \ \text{d}\eta.  \label{firstflat}
\end{equation} 

In accordance with Laplace's method, we rewrite the integral (\ref{firstflat}) as 

\begin{multline}
    \tdz^{-1} \sim \int_0^\varepsilon e^{-\frac{1}{4} \Pr \eta^2}\exp\left(-\frac{1}{2}\Pr\sum_{n=2}^\infty \frac{a_n}{n+1}\eta^{n+1}\right) \ \text{d}\eta \\ \sim \int_0^\varepsilon e^{-\frac{1}{4}\eta^2}\exp\left(\Pr\sum_{n=3}^\infty \bar{a}_n \eta^{n}\right) \ \text{d}\eta \label{bruhwhatimeanforreal}
\end{multline}

where we have defined 

\begin{subequations}
    \begin{align}
        \bar{a}_0 &= \bar{a}_1 = \bar{a}_2 = 0, \\ 
        \bar{a}_n &= -\frac{\displaystyle a_{n-1}}{2n} , n \geq 3.
    \end{align}
\end{subequations}

Next, we use the Taylor series of the exponential in (\ref{bruhwhatimeanforreal}) to obtain 

\begin{subequations}
\begin{align}
    \exp\left(\Pr\sum_{n=3}^\infty \bar{a}_n \eta^{n}\right) &= \sum_{n=0}^\infty \frac{Pr^n}{n!} \left\{\sum_{n=0}^\infty \bar{a}_n \eta^n \right\}^n \\ &= \sum_{n=0}^\infty \frac{\Pr^n}{n!}\sum_{k=3n}^\infty \lambda_{n, k}\eta^k,
\end{align}
\end{subequations}

where we have used the following
\begin{subequations}
    \begin{align}
        \left\{\sum_{j=0}^\infty \bar{a}_j \eta^j \right\}^n &= \sum_{k=0}^\infty \lambda_{n, k} \eta^k, \label{welpfirst} \\ 
        \lambda_{0, 0} &= 1, \hphantom{\sum\sum} \lambda_{0, k} = 0, k \geq 1, \\ \lambda_{n+1, k} &= \sum_{j=0}^k \lambda_{n, j} \bar{a}_{k-j}. 
    \end{align} \label{tn}
\end{subequations}

Eqs.~(\ref{tn}) are obtained by taking the powers of the relevant series recursively using Cauchy's product rule. Combining this expansion with (\ref{firstflat}) and noting that the leading term in the LHS of (\ref{welpfirst}) is of order $\eta^3$, $\lambda_{n, k} = 0$ when $k < 3n$, we obtain \begin{subequations} \begin{align}
    \tdz^{-1} &\sim \int_0^\varepsilon e^{-\frac{1}{4}\Pr \eta^2} \sum_{n=0}^\infty \frac{\Pr^n}{n!}\sum_{k=3n}^\infty \lambda_{n, k}\eta^k \ \text{d}\eta \label{one} \\ &\sim \sum_{n=0}^{\infty} \sum_{k=3n}^\infty \frac{1}{n!}\Pr^n \lambda_{n, k} \int_0^\varepsilon e^{-\frac{1}{4}\Pr \eta^2} \eta^k \ \text{d}\eta \label{two} \\ &\sim \sum_{n=0}^{\infty} \sum_{k=3n}^\infty \frac{1}{n!}\Pr^n \lambda_{n, k} \int_0^\infty e^{-\frac{1}{4}\Pr \eta^2} \eta^k \ \text{d}\eta \label{three} \\ &\sim \sum_{n=0}^\infty \sum_{k=3n}^\infty \frac{1}{n!}\Pr^n \lambda_{n, k} \Gamma\left(\frac{k+1}{2}\right)2^k \Pr^{(2n-k-1)/2} \label{four} \\ &\sim \sum_{\ell=1}^\infty \left[\sum_{n=0}^{\ell-1} \frac{1}{n!} \lambda_{n, \ell+2n-1} \Gamma\left(\frac{\ell+2n}{2}\right)2^{\ell+2n-1} \right] \Pr^{-\ell/2} ,\label{five}
\end{align} \label{bigflatcalc} \end{subequations}

yielding the result \begin{equation}
    \gamma_\ell = \sum_{n=0}^{\ell-1} \frac{1}{n!} \lambda_{n, \ell+2n-1} \Gamma\left(\frac{\ell+2n}{2}\right)2^{\ell+2n-1}. \label{flatans}
\end{equation}

In comparing (\ref{two}) with (\ref{three}), note that we have extended the upper bound on the integral from $\varepsilon$ to infinity, which only adds subdominant terms as $\Pr \to \infty$, as can be established with integration by parts. (\ref{five}) follows from (\ref{four}) by rewriting the sum for fixed order in $\Pr$. The resulting (\ref{flatans}) indeed produces exactly the values $\gamma_n$ from Appendix~\ref{GamTab}, while satisfying a slightly simpler recurrence relation than (\ref{gammarecforreal}), and makes clear that the coefficients $\gamma_n$ depend only on $\kappa$. 

\nocite{*}
\bibliography{aipsamp}

\end{document}